\documentclass[11pt]{article}
\usepackage[letterpaper,margin=20mm]{geometry}
\usepackage[utf8]{inputenc}
\usepackage{authblk}
\usepackage{amsmath}
\usepackage{amssymb}
\usepackage{graphicx}
\usepackage{cite}
\usepackage[english]{babel}

\title{Torsion or not torsion, that is the question}

\author{Yuri Bonder}
\affil{Instituto de Ciencias Nucleares, Universidad Nacional Aut\'onoma de M\'exico\\ Apartado Postal 70-543, Coyoac\'an, 04510, Cd. Mx., M\'exico\\ \textit{bonder@nucleares.unam.mx}}
\date{Essay written for the Gravity Research Foundation 2016 Awards for Essays on Gravitation}

\begin{document}

\maketitle

\begin{abstract}
A hypothesis of general relativity is that spacetime torsion vanishes identically. This assumption has no empirical support; in fact, a nonvanishing torsion is compatible with all the experimental tests of general relativity. The first part of this essay specifies the framework that is suitable to test the vanishing-torsion hypothesis, and an interesting relation with the gravitational degrees of freedom is suggested. In the second part, some original empirical tests are proposed based on the observation that torsion induces new interactions between different spin-polarized particles.
\end{abstract}

Any pseudo-Riemannian manifold admits a metric-compatible derivative operator with nonvanishing torsion. By definition, the failure of such an operator to commute, when applied to a scalar field, is related to the torsion tensor field ${T_{ab}}^c$. In addition, geometrically, torsion measures the failure of infinitesimal parallelograms to close \cite{Torsion}.

In general relativity (GR) it is assumed, without empirical support, that torsion vanishes identically. Of course, one may claim that the experimental success of GR justifies the vanishing-torsion hypothesis. However, as it is argued below, all GR tests are compatible with a nonvanishing torsion, and, as a basic assumption of the theory, it is paramount to experimentally test it. The main goal of this essay is to suggest alternative tests of the vanishing-torsion hypothesis, but before embarking on such a task, it is useful to specify the theoretical framework.

It is important to start by narrowing down the action that will be used. The experimental validation of GR allows one to conclude that the dominant term of the gravitational part of this action must be the Einstein--Hilbert action with the torsion-full derivative operator $\nabla_a$; this theory is known as the Einstein--Cartan theory and the modern formulation of such a theory, where spin plays a crucial role, can be traced back to Kibble \cite{Kibble} and Sciama \cite{Sciama}. (Ref.~\cite{Hehl} is a classic review on this topic.) Moreover, similarly to what happens in GR, the action can be examined under different dynamical approaches. One possibility is to impose that the metric-derivative vanishes; then torsion is a dynamically independent variable. The other alternative is to assume that the metric and the connection are independent degrees of freedom. Here the second approach is considered since, in such a scheme, it is not necessary to separate the connection \textit{a priori} into a metric and a torsion part, which seems somehow unnatural. Recall that, in standard general relativity (\textit{i.e.}, when torsion vanishes), the scheme in which the metric and the connection are independent is known as the Palatini approach \cite{Palatini}.

Since torsion is a property of the derivative operator, it should couple to matter through such an operator. It turns out that, with this restriction, torsion still couples (minimally) to spinors \cite{Hehl2}. This is the case considered in this essay since such particles are the best torsion-sensitive candidates to perform empirical tests. Of course, there are torsion theories where less restrictions are imposed, and which lead to a different phenomenology (for an example see Ref.~\cite{Ob}). Now, if the tetrad $e_\mu^a$ and spin connection $\omega_{a\mu\nu}$ are chosen as the independent dynamical variables that describe gravity (the conventions of Ref.~\cite{Wald} are used), then, the total action, for gravity and a spinor field $\Psi$ of mass $M$, takes the form
 \begin{eqnarray}
 S &=& \int d^4x e \left[R(e,\omega)+16\pi G \left(\frac{i}{2} e^a_\mu \bar{\Psi}\gamma^\mu \nabla_a\Psi -\frac{i}{2} e^a_\mu (\nabla_a\bar{\Psi})\gamma^\mu \Psi- M \bar{\Psi}\Psi\right) \right]\label{action}\\
 &=&\int d^4x e \left[R(e,\omega)+16\pi G \left(\frac{i}{2} e^a_\mu \bar{\Psi}\gamma^\mu \partial_a\Psi-\frac{i}{2} e^a_\mu (\partial_a\bar{\Psi})\gamma^\mu \Psi-\frac{1}{4} e^a_\mu \omega_{a\nu\rho} {\epsilon^{\mu\nu\rho}}_{\sigma} \bar{\Psi}\gamma_5 \gamma^\sigma \Psi - M\bar{\Psi}\Psi\right) \right],\nonumber
 \end{eqnarray}
where $e$ is determinant of the tetrad components, $R$ is the curvature scalar, and $G$ is Newton's constant. Also, for the second identity the explicit form of the covariant derivative acting on spinors is used \cite{Weinberg}, and $\epsilon_{\mu\nu\rho\sigma}$ are the components of the volume form in the dual tetrad basis, $\gamma^\mu$ are the Dirac matrices, $\gamma_5 =i \gamma^0\gamma^1\gamma^2\gamma^3$, and $\bar{\Psi}=\Psi^\dagger \gamma^0$. Note that the action (\ref{action}) has no free parameters and that the axial current associated with $\Psi$, which is given by $\bar{\Psi}\gamma_5 \gamma^\mu \Psi$, plays an important role. In addition, the spinor action explicitly depends on the derivative operator through $\omega_{a\mu\nu}$. Furthermore, bear in mind that spinors, which are fundamentally quantum-mechanical objects, are introduced through a semi-classical framework. Thus, for practical purposes, such particles behave as extended objects which can be spatially overlapped.

The variation of the action (\ref{action}) with respect to the tetrad gives an equation whose symmetric part reduces to Einstein equations and the antisymmetric part reflects the freedom to perform local Lorentz transformations (in the presence of torsion, the Ricci tensor is not necessarily symmetric). On the other hand, the spin-connection variation yields
\begin{equation}\label{spin connection eom}
 0= 2\nabla_a(e_\mu^{[a} e^{c]}_\nu) - 4 \delta^{\rho}_{[\mu} e^{[a}_{\nu]} e^{c]\sigma} \omega_{a\rho\sigma} - e^{ a}_\mu e^{ b}_\nu {T_{ab}}^c + 4\pi G e^c_\rho {\epsilon^\rho}_{\mu\nu\sigma} j_5^\sigma,\\
\end{equation}
where $j_5^\mu$ is the total axial current. The solution of this equation is $\omega_{a\mu\nu} = e_{\mu b} \nabla_a e^b_\nu$ and
\begin{equation}\label{torsion}
 {T_{ab}}^c = 4\pi G e^d_\mu {\epsilon^c}_{abd} j_5^\mu.
 \end{equation}
Observe that torsion is linked to $j_5^\mu$ through an algebraic equation, which implies that torsion does not propagate. In addition, $j_5^\mu$ has contributions of all the spinor fields under consideration. Moreover, since the axial current is generated by spin-polarized spinors, to look for torsion in the present context, the experiments need to be done inside a spin-polarized medium. In this sense the claim that all GR tests are compatible with a nonvanishing torsion holds, since, clearly, these tests are done in situations where, at least on average, $j_5^\mu=0$.

In the standard (torsionless) general relativity in the presence of spinors, when the gravitational degrees of freedom are treated \textit{\`a la} Palatini, the solution to the equation of motion associated with the spin connection is $\omega_{a\mu\nu} =e_{\mu b} \nabla_a e^b_\nu - 2\pi G e^\rho_{ a} {\epsilon}_{\rho\mu\nu\sigma} j_5^\sigma$. Interestingly, the axial current term is not present when one takes the `standard' approach, namely, when the connection is fixed \textit{a priori} by the metric-compatibility condition. This is a particular example of the fact that these formalisms are physically inequivalent when the matter action depends on the connection \cite{Tsamparlis}. In any case, this result suggests that the issue of whether torsion vanishes could be closely related to the question of whether gravity is described by the standard or the Palatini approach.

Recall that main purpose of this essay is to suggest alternative experimental strategies to test for torsion. It should be mentioned that some consequences of torsion have been explored, for example, in Refs.~ \cite{consequences2,consequences3,consequences4,consequences5}, and reviews including broad discussions on this topic may be found in Refs.~\cite{Hammond,Shapiro,consequences1}. To analyze the experimental consequences of torsion in the context outlined above, it is useful to appeal to a test-particle approximation where the test spinor is denoted by $\psi$ and all other spinor fields, which are collectively called source spinors, are assumed to generate the axial current $J_5^\mu$. The test-spinor equation of motion, assuming it has mass $m$, takes the form
\begin{eqnarray}\label{Dirac}
 0&=& i e^a_\mu\gamma^\mu \partial_a \psi + \frac{i}{2}(\nabla_a e^a_\mu)\gamma^\mu \psi-\frac{1}{4} e^a_\mu \omega_{a\rho\sigma} {\epsilon^{\mu\rho\sigma}}_\nu \gamma_5 \gamma^\nu \psi- m\psi.
\end{eqnarray}
Note that torsion is present in equation (\ref{Dirac}) through $\nabla_a$ and $\omega_{a\mu\nu}$. Therefore, it is possible to use equation (\ref{torsion}) to replace ${T_{ab}}^c$ in equation (\ref{Dirac}), leading to the well-known nonlinear terms in $\psi$ \cite{selfint}, and additional interactions between the test and source spinors.

For the purpose of characterizing the torsion interactions, it is useful to obtain the Hamiltonian. As it is customary, this Hamiltonian can be read off from equation (\ref{Dirac}). For particular experimental situations where it is possible to neglect the torsion self-interactions, the special-relativistic corrections, and the curvature effects, the Hamiltonian, which is obtained using standard methods \cite{FW}, takes the form 
\begin{equation}\label{nonrel Hamiltonian}
H = m+\frac{\vec{p}^2}{2m} - 3 \pi G \vec{J}_5 \cdot \vec{\sigma} + \frac{3\pi G}{2m} \vec{\sigma}\cdot \left(\vec{p} J_5^0 + J_5^0 \vec{p}\right),
\end{equation}
where $\vec{p}$ is the momentum operator, $\sigma^i$ are the Pauli matrices, and the standard $3$-dimensional vectorial notation is utilized. The first two terms in this Hamiltonian are the conventional rest and kinetic energy terms, and the last two terms are due to torsion and are suppressed by $G$. The fact that all torsion effects are highly suppressed is well known, in fact, in the case of nucleons, the length scale associated to torsion effects is $10^{-28} \ {\rm m}$ \cite{Hehl2}, indicating that its experimental search will be daunting. Moreover, from the explicit form of these torsion interactions, it is evident that, to produce nontrivial contributions, the test spinor has to be in a spin-polarized state and overlapping the source spinors.

The $\vec{J}_{5} \cdot \vec{\sigma}$ term in the Hamiltonian (\ref{nonrel Hamiltonian}) can be probed in polarized-neutron transmission experiments through polarized media \cite{MikePersonal}. Three alternatives look promising in this direction: First, there is an experiment where a polarized neutron beam is sent into a polarized Holmium target \cite{holmium}, which is, in principle, sensitive to torsion effects. However, in this experiment, the torsion signal would behave like the experimental noise, which is discarded. Second, to perform a neutron spin-rotation experiment, as in Ref.~\cite{Mike}, but where the target is spin-polarized liquid Helium. Third, there are spin sources that have an extremely high spin-polarization density but are insensitive to magnetic effects \cite{Eotwash}, and it is conceivable to perform an experiment where a polarized neutron beam passes through such a spin source. Interestingly, one could separate torsion effects from spurious signals by, for example, changing the relative spin orientation, or using that, in contrast to the electromagnetic interaction, this torsion term is momentum independent. In addition, in this case one would probe a neutron-electron coupling where the noise is basically electromagnetic. This is a huge improvement with respect to the other two proposals where the noise is due to the strong interaction. It should be also stressed that the present-day methods for generating and manipulating large flows of ultracold neutrons \cite{UCN} may help reduce the statistical errors.

Interferometry tests are sensitive to the last term in the Hamiltonian (\ref{nonrel Hamiltonian}). In fact, it can be shown that the phase difference produced by such a term is related to the difference of the line integral of $J_5^0$ along the interferometer arms. Naively, it is tempting to eliminate the noise from spin-spin interactions by surrounding the spin-polarized region as in an Aharonov--Bohm setup. However, $J_5^0$ is not a potential, thus, no phase difference is produced if $J_5^0$ vanishes in the integration region. Hence, to look for these effects one must do interferometry inside a spin-polarized medium, which is technologically challenging. In any case, looking for the effects of both unconventional terms in the Hamiltonian (\ref{nonrel Hamiltonian}) may be a valuable tool to separate possible torsion effects from other interactions.

In conclusion, the vanishing-torsion hypothesis of GR has no empirical support and it must be empirically tested. Here some original strategies to test this hypothesis are suggested, whose ultimate sensitivities and consequences are impossible to foresee. Yet, these tests should allow us to quantify the validity of the vanishing-torsion hypothesis, and a potential discovery of a nonvanishing spacetime torsion would have deep conceptual consequences.

\section*{Acknowledgement}
I thank the referee for his/her useful comments and Mike Snow for his experimental suggestions. This work was done with financial support from UNAM-DGAPA-PAPIIT Project No. IA101116.

\end{document}